# Observation of normal-force-independent superlubricity in mesoscopic graphite contacts

Cuong Cao Vu, [1,2] Shoumo Zhang, [2,3] Michael Urbakh, [4] Qunyang Li, [2,3,5] Q.-C. He, [1,6,]* and Quanshui Zheng, [2,3,5,]*

[1] School of Mechanical Engineering, Southwest Jiaotong University, Chengdu, 610031, China

[2] Center for Nano and Micro Mechanics, Tsinghua University, Beijing 100084, China

[3] Department of Engineering Mechanics, Tsinghua University, Beijing 100084, China

[4] School of Chemistry, Tel Aviv University, 69978 Tel Aviv, Israel

[5] Applied Mechanics Laboratory and State Key Laboratory of Tribology, Tsinghua University, Beijing 100084, China

[6] Université Paris-Est, Laboratoire de Modélisation et Simulation Multi Echelle, UMR 8208 CNRS, 5 bd Descartes, 77454 Marne-la-Vallée, France

* Corresponding authors: zhengqs@tsinghua.edu.cn, qi-chang.he@u-pem.fr

**Abstract:**

We investigate the dependence of friction forces on normal load in incommensurate micrometer size contacts between atomically smooth single-crystal graphite surfaces under ambient conditions. Our experimental results show that these contacts exhibit superlubricity (super-low friction), which is robust against the application of normal load. The measured friction coefficients are essentially zero and independent of the external normal load up to the maximum pressure of our experiment, 1.67 MPa. The observation of load-independent superlubricity in micro-scale contacts is a promising result for numerous practical applications.

Friction and wear impose serious constraints and limitations on the performance and lifetime of micro-machines and, undoubtedly, will impose even more severe constraints on the emerging technology of nano-machines [1-5]. Using materials that possess superlow friction may provide a desired route to overcome these major problems. A mechanism for superlow dry friction, which arises from the structural incompatibility of two contacting solids, was first suggested by Hirano and Shinjo [6,7]. This phenomenon is also referred to as superlubricity. Recent experimental observations of superlubricity for graphite samples of micrometer size under ambient conditions [8,9] and at extremely high speed (up to 25 m/s) [10], and superlubricity for centimetre-long carbon nanotubes [11] are promising results for many practical applications.

The question of largest interest for fundamental studies of friction and numerous applications is how robust is the superlubric state against the normal load. For nanoscale incommensurate contacts between graphene flakes and graphite surfaces a sudden increase of friction with increasing load has been observed in experiments and simulations [12]. This loading effect has been attributed to the load-induced locking of the flakes as a result of vertical motion of edge atoms. A breakdown of superlubricty can also occur if the application of normal load leads to strong in-plane distortions resulting in local commensurability of contacting surfaces [13,14]. On the other hand, so far all experimental studies of superlubrictity at micro- and macroscales [8-10,15] have been performed without application of normal load, and we still do not know how the load will affect the superlubric state at the scales relevant to tribological applications. Here, we investigate this fundamental question measuring friction between microscopic graphite mesas in ambient conditions.

Our work is based on the recent discoveries of self-retraction motion (SRM) and superlubricity in graphite-graphite contacts [8,15]. Square graphite mesas of the sizes 3 $\mu$m × 3 $\mu$m and height 1 $\mu$m with a $SiO_2$ cap of thickness 200 nm on each mesa are fabricated similar to that described in the references [10,15-18] and in the Supplementary Material (SM) [19]. We use an optical microscope (OM, HiRox KH-3000) and a micro-manipulator mm3A (Kleindiek MM3A) to select mesas that exhibit the SRM property. The chosen samples were used in subsequent atomic force microscope (AFM, NT-MDT) experiments performed at various normal loads. As illustrated in Fig. 1(a), the SRM phenomena are observed when the top graphite layer is sheared laterally relative to those below and slides back when the layer is released, retracting towards its original, overlapping position. It was shown that the SRM occurs only for the incommensurate contacts between the top and bottom flakes with atomically flat [0001] graphite surfaces, and in these contacts the interfacial friction is nearly zero (superlubricity) [8]. The superlubric interface remains clean during the sliding experiments since the SRM removes contaminants adsorbed on the exposed [0001] graphite surfaces, providing the self-cleaning (nanoeraser) effect [20,21].

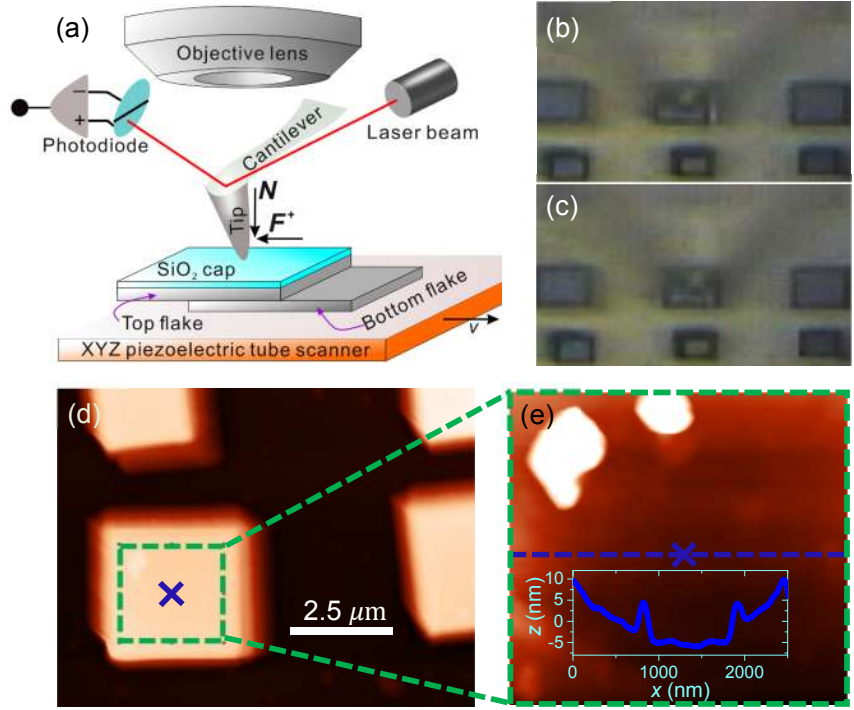

FIG. 1 (color online). Schematic diagram of the experimental setup. (a) An objective lens is coupled to the AFM head to observe the relative movement of the sheared flake with respect to the substrate. Experiments are performed in an ambient condition with 1.5 $\mu m/s$ of sliding velocity ($v$) and 0.8 $\mu m$ of sliding distance ($x$). The normal ($N$) and lateral ($F$) forces are applied to the top flake through its $SiO_2$ cap using the same AFM probe that contacts the central area of the cap (the cross in (d)). The two force components are measured simultaneously by the AFM. (b,c) Two optical images obtained for the top flake sheared in the forward and backward directions, respectively. (d,e) AFM topographic images of the selected sample. In order to generate sufficiently large lateral forces for very low normal loads, we made an indentation in the central area of the $SiO_2$ cover. The inset to the panel (d) shows the AFM scanning profile (the blue curve) measured along the dashed blue line across the indentation, see (d). The profile shows that the width and depth of the indentation are about 1 $\mu m$ and 10 nm, respectively.

As illustrated in Fig. 1(a), our experimental setup includes a commercial Ntegra upright AFM (NT-MDT, Russia), a 100 $\mu m$ XYZ piezoelectric tube scanner, and a high numerical aperture objective lens (× 100, Mitutoyu, Japan), equipped with a special AFM cantilever (VIT-P tip, NT-MDT, Russia, the nominal spring constant is 50 N/m) so that the AFM tip is visible when it is in contact with the test graphite mesa samples. The sample is fixed on the scanner and is pressed with the AFM tip acting at the indentation made in advance (see Figs. 1(d), 1(e), or the SM [19]) in the central area of the $SiO_2$ cap. To avoid a complex analysis of the effect of the top flake rotation on friction, we report results that correspond to SRM without a notable rotation, see the SM [19]. The normal force ($N$) applied by the tip to the SiO$_2$ cap at the top of graphite flake can be precisely measured (with an accuracy of 4%, see the SM [19]) and controlled through the AFM feedback system. Operating the scanner with a normal force $N$ allows to induce a relative lateral motion between the top and bottom graphite flakes (Fig. 1(a)). The corresponding lateral (shear) force ($F$), which is applied to the cap by the same tip, can be also precisely measured (with the resolution of the order of 0.5 nN and an accuracy of 0.7%, see the SM [19]) by the AFM. A standard AFM calibration method [22-24] is adopted for both the lateral and normal force measurements, as detailed in the SM [19]. Experiments are carried out in an ambient condition (temperature ~ 25°C, relative humidity ~ 30%) in Beijing.

Figure 2(a) presents a typical trace showing both forward ($F^+$, the red curve) and backward ($F^-$, the blue curve) lateral force for a fixed normal load ($N$ = 3.72 $\mu$N, see the left inset to Fig. 2(a)). The sliding speed ($v$) is 1.5 $\mu$m/s and the maximum lateral displacement ($x_{max}$) is 800 nm.

The force-displacement loops have an excellent repeatability for 500 cycles (the maximum cycle number which has been tested). The results for five cycles are shown in the right inset to Fig. 2(a). It should be noted that during the shearing/retracting motion under the ambient condition the exposed surfaces adsorb contaminants. If the adsorbed contaminants, which are located in front of the top flake, would enter into the contact area over the next sliding/retracting motions, then the subsequent force loops will be significantly different from the previous ones, and the repeatability will be lost [25]. This suggests that the SRM removes contaminants adsorbed on the exposed graphite surfaces [20,21].

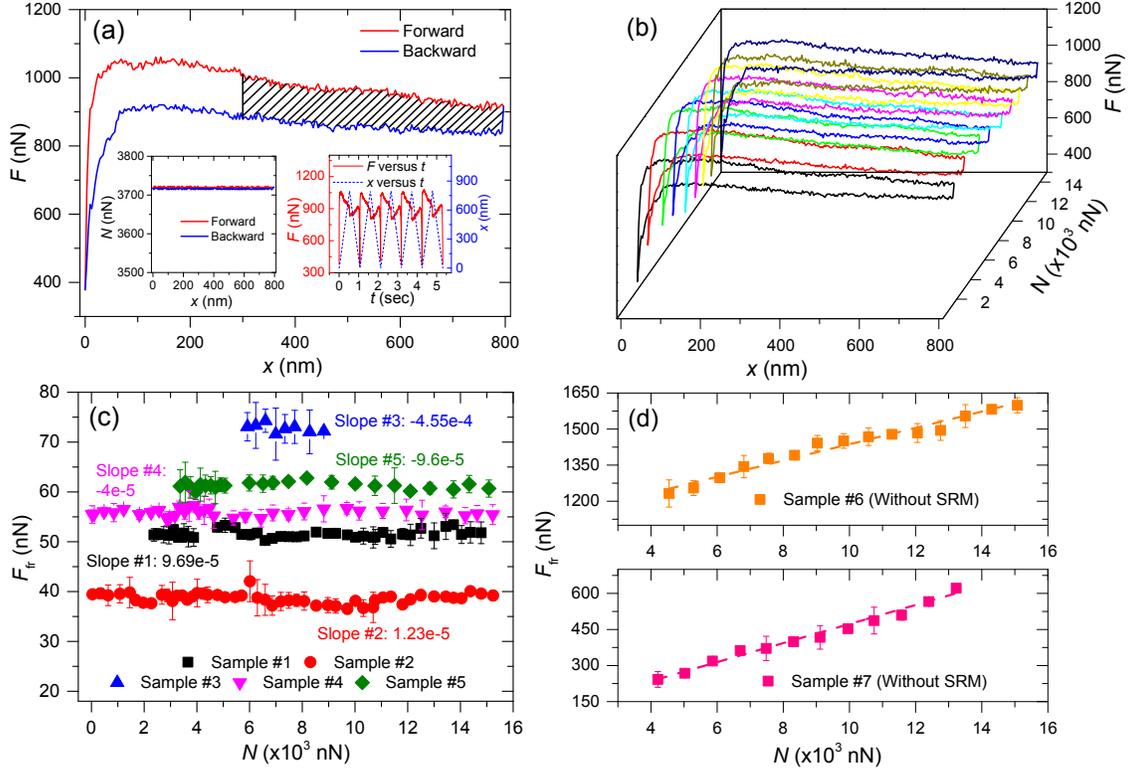

FIG. 2 (color online). The lateral force ($F$) and applied normal force ($N$) for a typical mesa sample of $3\ \mu m \times 3\ \mu m$. (a) Lateral force's loop corresponds to forward ($F^+$, red) and backward ($F^-$, blue) directions. The corresponding normal forces are shown in the left inset. The right insert illustrates the repeatability of the force loops for five cycles. (b) The lateral force – displacement loops measured for various normal forces $N$ up to 15 $\mu N$. (c) The dependencies of the friction forces on the normal force measured for five self-retractable mesas of the sizes $3\ \mu m \times 3\ \mu m$. The friction force ($F_{fr}$) is estimated as the mean value of the $\frac{1}{2}(F^+ - F^-)$ over the sliding displacement range 300 nm $< x <$ 800 nm. (d) Variations of friction forces with normal load measured for two randomly chosen graphite mesas of the same size $3\ \mu m \times 3\ \mu m$ but not showing SRM. Compared with the $N$-independent friction observed for the superlubric contacts, as shown in (c), here we found usual linear increase of friction with $N$, and the friction forces are one order of magnitude higher than for the superlubric state.

Figure 2(b) shows a typical set of lateral force loops measured for different values of normal force $N$. A remarkable feature observed from Fig. 2(b) is that the shapes and areas of all these loops are essentially unchanged by $N$.

It should be noted that the lateral force loops observed in our experiments, which are shown in Figs. 2(a), 2(b) differ significantly from the conventional friction loops measured in AFM experiments [26]. They show two "unusual" features. First, the lateral forces are always positive. Second, the differences between the lateral forces measured during forward and backward motion decrease with increasing the displacement, $x$.

The reason for the observed features can be easily understood taking into account that the forces measured during the sliding motion in the forward and backward directions can be written as $F^{\pm} = F_{\text{ret}} \pm F_{\text{fr}}^{\pm}$, where $F_{\text{ret}} = \Gamma L$ is the retraction force that drives the self-retraction motion [15] with $\Gamma$ (= 0.37 J/m²) being the graphite cleavage energy [9,18] and $L$ (= 3 μm) is the mesa size, and $F_{\text{fr}}^{+}$ and $F_{\text{fr}}^{-}$ are the friction forces experienced by the flake during the forward and backward motions, respectively. The retraction force gives the main contribution to the measured forces, and as a result $F^{\pm} > 0$. The increase of displacement, $x$, leads to the reduction of the contact area and the edge lengths, and as a result the friction forces, $F_{\text{fr}}^{\pm}$, and the measured difference, $F^{+} - F^{-}$, decreases with $x$. In addition, the friction forces $F_{\text{fr}}^{+}$ and $F_{\text{fr}}^{-}$ could be different because of adsorption on the graphite surfaces exposed to air during the shearing time.

Friction forces in incommensurate contacts scale with the contact area $A$ as $F_{\text{fr}}^{\pm} = F_0^{\pm}(N) A^{\gamma}$, where the exponent $\gamma \leq 0.5$ [9,27-29]. It should be noted that in our experiments the area of contact between the relatively sliding top and the bottom flakes changes during the shearing, as $A = L(L - x)$. Then, integrating the force loops in Fig. 2(a) over the displacement range $x_{\min} = 300$ nm $< x <$ $x_{\max} = 800$ nm, where the forces $F^{\pm}$ are nearly constant, we get

$$\frac{1}{2}\langle F^+(x) - F^-(x)\rangle = \frac{1}{2(x_{\max} - x_{\min})} \int_{x_{\min}}^{x_{\max}} [F^+(x) - F^-(x)]dx$$

$$= \frac{F_0^+(N) + F_0^-(N)}{2} A^\gamma \left[\frac{\left(1 - \frac{x_{\min}}{L}\right)^{1+\frac{\gamma}{2}} - \left(1 - \frac{x_{\max}}{L}\right)^{1+\frac{\gamma}{2}}}{\left(1 + \frac{\gamma}{2}\right)\frac{x_{\max} - x_{\min}}{L}}\right] \quad (1)$$

$$\approx \frac{F_0^+(N) + F_0^-(N)}{2} A^\gamma$$

The last approximation in Eq. (1) is obtained for small $\frac{x_{\max}}{L}$ that is indeed the case in our experiments. Thus, measuring $\frac{1}{2}\langle F^+(x) - F^-(x)\rangle$ we define the mean friction force,

$$\langle F_{\mathrm{fr}}\rangle = \frac{1}{2}\langle F_{\mathrm{fr}}^+ + F_{\mathrm{fr}}^-\rangle = \frac{F_0^+(N) + F_0^-(N)}{2} A^\gamma \quad (2)$$

Figure 2(c) presents the dependencies of the mean friction force, $\langle F_{\mathrm{fr}}\rangle$, on the normal force, $N$, which have been measured for five samples. For all samples we find extremely small and even negative values of the mean slope of $\langle F_{\mathrm{fr}}\rangle$ versus $N$ lying in the range between $9.69 \times 10^{-5}$ and $-4.55 \times 10^{-4}$. Similar slopes of the friction force are observed for more than a dozen of mesas exhibiting SRM. Thus, within the accuracy of our measurements (see error bars in Fig. 2(c)) the friction coefficient (the slope of $\langle F_{\mathrm{fr}}\rangle$ vs $N$) in incommensurate micrometer size contacts between atomically smooth single-crystal graphite surfaces is essentially zero and independent of the normal load, up to a normal pressure of 1.67 MPa. In comparison, for the graphite flakes, which do not exhibit the SRM, we found a significant increase of friction forces with the increasing load, where the mean slope of $\langle F_{\mathrm{fr}}\rangle$ versus $N$ is about 0.03 (see for, example, results in Fig. 2(d)).

Recently, the cleavage energy $\Gamma$ of graphite was directly measured independently by two research groups [9,18]. Both works were based on the same SRM phenomenon, but used the samples of different sizes and different techniques: in ref. [9] the 100~400 nm sized graphite mesas have been studied with AFM, whereas in ref. [18] a force sensor has been employed to perform measurements with the 3~4 $\mu$m sized graphite mesas. The experiments gave different values of $\Gamma$, 0.227±0.005 J/m$^2$ and 0.37±0.01 J/m$^2$, respectively. We note that the estimation of $\Gamma$ using the relation $\Gamma = F^+/L$ with $L$=3 $\mu$m and the force measured by AFM in the present study gives $\Gamma$ of about 0.35 J/m$^2$. Regarding the shear stress in the incommensurate graphite contacts at zero normal load, our result, ~ 0.057 MPa, is very close to the value reported by Koren et al. [9], ~ 0.051 MPa, but significant smaller than the value reported by Dienwiebel et al. [26], that is, ~ 6.1 MPa.

The micrometer size frictional contacts considered in this study are too larger to be studied using molecular dynamics simulations (MD). However, the insight into the mechanism of super-low friction observed here can be revealed through the simplified theoretical models such as the Frenkel-Kontorova-Tomlinson model [13,30], and from the simulations performed for small systems [12,13]. The previous theoretical results and simulations [12,13,31,32] demonstrated that the superlubric regime of friction with close to zero friction coefficient, which is observed here for incommensurate contacts, up to a normal pressure of 1.67 MPa, corresponds to smooth sliding modulated by the periodicities of contacting surfaces. The increase of normal load enhances the potential corrugation at the frictional interface, and above a critical load, for which the stiffness of interfacial potential corrugation reaches the intralayer stiffness of contacting materials, the transition from smooth sliding to stick-slip and the corresponding increase of the friction coefficient should occur. This breakdown of superlubricity is caused by the load-induced rearrangement of the surface atoms, making the interface configuration locally commensurate. The local commensuration may be the most energetically favourable near the edges of the substrates [12]. MD simulations performed for nanoscale graphite flakes on the graphite surface [12] found that the critical normal force corresponding to the transition to high-dissipative stick-slip regime is of the order 0.6 nN per the flake atom. This force corresponds to the normal pressure of ~ 23 GPa, which is significantly higher than the maximum pressure in our experiments, 1.67 MPa. The latter value is close to the maximum load that can be reached by the special AFM probe used in our measurements.

Figure 2(c) shows a notable difference between the friction forces observed for five samples, which have been prepared using the same fabrication procedure and have the same size (3 $\mu$m × 3 $\mu$m). Intuitively, one would expect a certain variation of friction forces with degree of contamination located in particular at the edges of the flakes. To test this hypothesis, we measured friction forces for these five samples according to the following protocols: (1) before heating, (2) after heating the samples at 150°C for one hour and then exposing them to ambient conditions and room temperature for about 10~30 minutes (the minimum time needed to prepare the experiment), (3) after exposing the samples to ambient conditions for 14 days, and (4) after reheating the samples at 200°C for one hour and then exposing them to ambient conditions and room temperature for about 10~30 minutes. The results show in Fig. 3 demonstrate that large scattering of friction forces observed for samples exposed to ambient conditions and room temperature for an extended time (square navy and diamond purple symbols) is removed after a heating at 150°C or 200°C (up triangle pink and right triangle violet symbols). This allows to conclude that the scattering of the results in Fig. 2 is indeed the effect of contaminants on the friction force, which diminishes after the heating processes. Thus, our experiments demonstrate that the incommensurate micrometer size contacts between atomically smooth single-crystal graphite surfaces provide a super-low friction force (superlubricity), which is independent of normal load.

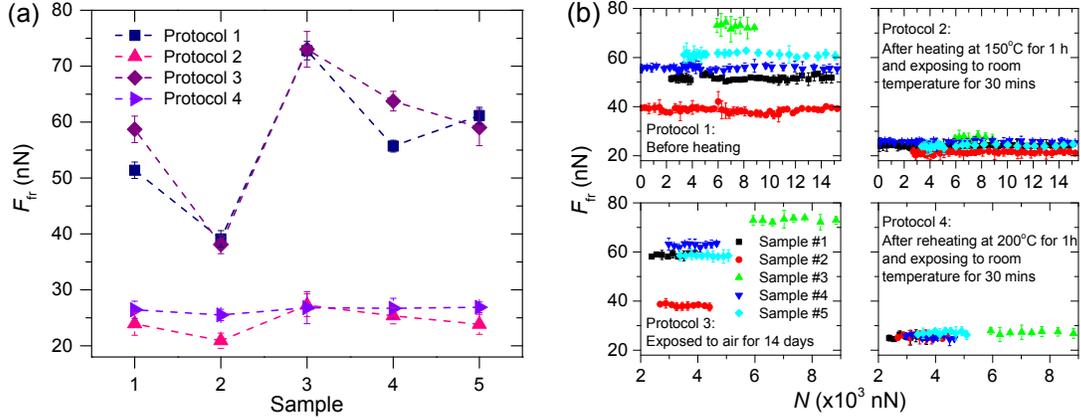

FIG. 3 (color online). Effect of heating on friction forces measured with the five selected mesas. (a) Friction forces measured at room condition according to four different protocols: (1) before heating (squares, navy color), (2) after heating at 150°C for 1 hour and then exposing to room temperature for 10~30 minutes (up triangle pink color), (3) exposed to air for 14 days (diamonds, purple color), and (4) after reheating at 200°C for one hour and then exposing to room temperature for 30 minutes (right triangle violet). (b) Variation of friction forces with gradually increasing normal force observed for the samples, which have been treated according to the protocols 1-4, respectively. The error bars show the standard deviation of measurements for each plot.

The contribution of contaminants to friction, $F$, is the key but still open question in superlubricity studies. The observation of "normal-force-independence" of friction force in the superlubric state allows to draw important conclusions on the contribution of contaminants to friction, as explained below. The contribution of contaminants to the friction force can be separated into two terms: area, $F_{fr}^{(area)}$, and edge, $F_{fr}^{(edge)}$, contributions. The first one comes from the contaminants located at the contact area between the top and bottom graphite flakes, whereas the second one comes from the contaminants located at the edges of the driven flake. According to the theoretical works [25,33], the term $F_{fr}^{(area)}$ follows the Amontons-Coulomb law, $F_{fr}^{(area)} = \mu(\rho)N$, with the friction coefficient $\mu(\rho)$ depending on the surface density of contaminants. The coefficient $\mu(\rho)$ was estimated as $\sim 10^{-1}$ for $\rho \sim 10^{-1}$, where $\rho$ is defined as a number of absorbed atoms per surface atom. At low values of $\rho$ a linear dependence of $\mu(\rho)$ on $\rho$ is expected, and we have $\mu(\rho) \sim \rho$. Previous experiments show that, when freshly prepared graphite surface is exposed to ambient condition, it is fully covered by atomic contaminants within time of the order of minute [34]. Taking into account that the duration of each loading-unloading loop is 1s, the surface coverage can be estimated as $\rho \sim 10^{-2}$. Then, assuming that the main contribution to friction comes from the contaminants located at the contact area between the flakes, we get $\mu(\rho) \sim 10^{-2}$, which is two orders of magnitude higher than the measured friction coefficient, $\mu(\rho) < 10^{-4}$. This indicates that most of contaminants (~99%) adsorbed at the exposed area are swept away by the edge of the moving flake in each loading-unloading loop and do not enter the contact area. Thus, the main contribution to friction is given by the edges, $F_{\text{fr}}^{(\text{edge})}$. The results presented in Fig.3 and in Fig. 2 in ref. [18] show that $F_{\text{fr}}^{(\text{edge})}$ can be reduced by one order of magnitude after heating the samples at 150-200°C. Ability to remove the contaminants by heating at 150°C indicates that they are caused by a weak adsorption through, for instance, hydrogen bonds.

To conclude, we investigated a variation of friction force with increasing normal load in incommensurate contacts between atomically smooth single-crystal graphite surfaces. Our measurements demonstrate that superlubricity is robust against the application of normal load up to 1.67 MPa, corresponding to a normal load of 15 $\mu$N acted on a square flake of 3 $\mu$m × 3 $\mu$m. In the investigated range of normal load for all selected samples the friction coefficient remains smaller than $3 \times 10^{-5}$, and after unloading the graphite flakes still exhibit SRM. The observation of load-independent superlubricity in micro-scale contacts is a promising result for practical applications, such as MEMS, sensors, fast switches and more.

An open challenge is to experimentally investigate the load-induced transition from superlubric to highly dissipative stick-slip motion. The maximum load allowed by our AFM probe is 15 $\mu$N that is four orders of magnitude lower than the prediction of molecular dynamics simulations [12] for the critical load corresponding to this transition. This predicted pressure is much higher than the current experimental capabilities.

**Acknowledgement.** C.C.V. thanks to Dr. Jing Li from Chinese Academy of Sciences and Dr. Lin Ge from NT-MDT Beijing Office for their helps in experimental device support, to Prof. Xing Yang from Tsinghua University for his valuable discussions during the preparation of this manuscript, to Mr. C.Y. Qu from Q.S.Z.'s group for his help in calibration. Q.S.Z., Q.C.H., and Q.Y.L. acknowledge the financial support from the National 973 Project through Grant No. 2013CB934201, the Cyrus Tang Foundation through Grant No. 202003, the National Natural Science Foundation of China (11272177, 11422218), the Beijing Municipal Science & Technology Commission through Program No. Z151100003315008, and the Tsinghua University Initiative Scientific Research through Program No. 2012Z01015, 2012Z02134. Q.S.Z. and M.U. acknowledge the support of XIN center, Tsinghua-Tel Aviv Universities.